\documentstyle[12pt,epsf,rotate]{article}
\begin{document}
\baselineskip 16pt plus 2pt minus 2pt

\def\bra#1{{\langle#1\vert}}
\def\ket#1{{\vert#1\rangle}}
\def\sst#1{{\scriptscriptstyle #1}}
\def\rbra#1{{\langle #1 \vert\!\vert}}
\def\rket#1{{\vert\!\vert #1\rangle}}
\def\simge{\hspace*{0.2em}\raisebox{0.5ex}{$>$}
     \hspace{-0.8em}\raisebox{-0.3em}{$\sim$}\hspace*{0.2em}}
\def\simle{\hspace*{0.2em}\raisebox{0.5ex}{$<$}
     \hspace{-0.8em}\raisebox{-0.3em}{$\sim$}\hspace*{0.2em}}
\def\psibar{{\bar\psi}}
\def\Gmu{{G_\mu}}
\def\alr{{A_\sst{LR}}}
\def\wpv{{W^\sst{PV}}}
\def\evec{{\vec e}}
\def\notq{{\not\! q}}
\def\notk{{\not\! k}}
\def\notp{{\not\! p}}
\def\notpp{{\not\! p'}}
\def\notder{{\not\! \partial}}
\def\notcder{{\not\!\! D}}
\def\notA{{\not\!\! A}}
\def\notv{{\not\!\! v}}
\def\Jem{{J_\mu^{em}}}
\def\Jana{{J_{\mu 5}^{anapole}}}
\def\nue{{\nu_e}}
\def\mn{{m_\sst{N}}}
\def\mns{{m^2_\sst{N}}}
\def\me{{m_e}}
\def\mes{{m^2_e}}
\def\mq{{m_q}}
\def\mqs{{m_q^2}}
\def\mz{{M_\sst{Z}}}
\def\mzs{{M^2_\sst{Z}}}
\def\ubar{{\bar u}}
\def\dbar{{\bar d}}
\def\sbar{{\bar s}}
\def\qbar{{\bar q}}
\def\sstw{{\sin^2\theta_\sst{W}}}
\def\gv{{g_\sst{V}}}
\def\ga{{g_\sst{A}}}
\def\pv{{\vec p}}
\def\pvs{{{\vec p}^{\>2}}}
\def\ppv{{{\vec p}^{\>\prime}}}
\def\ppvs{{{\vec p}^{\>\prime\>2}}}
\def\qv{{\vec q}}
\def\qvs{{{\vec q}^{\>2}}}
\def\xv{{\vec x}}
\def\xpv{{{\vec x}^{\>\prime}}}
\def\yv{{\vec y}}
\def\tauv{{\vec\tau}}
\def\sigv{{\vec\sigma}}
\def\sst#1{{\scriptscriptstyle #1}}
\def\gpnn{{g_{\sst{NN}\pi}}}
\def\grnn{{g_{\sst{NN}\rho}}}
\def\gnnm{{g_\sst{NNM}}}
\def\hnnm{{h_\sst{NNM}}}
\def\gvnn{{g_\sst{NNV}}}
\def\gonn{{g_{\sst{NN}\omega}}}
\def\fonn{{f_{\sst{NN}\omega}}}

\def\xivz{{\xi_\sst{V}^{(0)}}}
\def\xivt{{\xi_\sst{V}^{(3)}}}
\def\xive{{\xi_\sst{V}^{(8)}}}
\def\xiaz{{\xi_\sst{A}^{(0)}}}
\def\xiat{{\xi_\sst{A}^{(3)}}}
\def\xiae{{\xi_\sst{A}^{(8)}}}
\def\xivtez{{\xi_\sst{V}^{I=0}}}
\def\xivteo{{\xi_\sst{V}^{I=1}}}
\def\xiatez{{\xi_\sst{A}^{I=0}}}
\def\xiateo{{\xi_\sst{A}^{I=1}}}
\def\xiva{{\xi_\sst{V,A}}}

\def\rvz{{R_\sst{V}^{(0)}}}
\def\rvt{{R_\sst{V}^{(3)}}}
\def\rve{{R_\sst{V}^{(8)}}}
\def\raz{{R_\sst{A}^{(0)}}}
\def\rat{{R_\sst{A}^{(3)}}}
\def\rae{{R_\sst{A}^{(8)}}}
\def\rvtez{{R_\sst{V}^{I=0}}}
\def\rvteo{{R_\sst{V}^{I=1}}}
\def\ratez{{R_\sst{A}^{I=0}}}
\def\rateo{{R_\sst{A}^{I=1}}}

\def\mro{{m_\rho}}
\def\mks{{m_\sst{K}^2}}
\def\mpi{{m_\pi}}
\def\mpis{{m_\pi^2}}
\def\mom{{m_\omega}}
\def\mphi{{m_\phi}}
\def\Qhat{{\hat Q}}

\def\FOS{{F_1^{(s)}}}
\def\FTS{{F_2^{(s)}}}
\def\GAS{{G_\sst{A}^{(s)}}}
\def\GES{{G_\sst{E}^{(s)}}}
\def\GMS{{G_\sst{M}^{(s)}}}
\def\GATEZ{{G_\sst{A}^{\sst{I}=0}}}
\def\GATEO{{G_\sst{A}^{\sst{I}=1}}}
\def\mdax{{M_\sst{A}}}
\def\mustr{{\mu_s}}
\def\rsstr{{r^2_s}}
\def\rhostr{{\rho_s}}
\def\GEG{{G_\sst{E}^\gamma}}
\def\GEZ{{G_\sst{E}^\sst{Z}}}
\def\GMG{{G_\sst{M}^\gamma}}
\def\GMZ{{G_\sst{M}^\sst{Z}}}
\def\GEn{{G_\sst{E}^n}}
\def\GEp{{G_\sst{E}^p}}
\def\GMn{{G_\sst{M}^n}}
\def\GMp{{G_\sst{M}^p}}
\def\GAp{{G_\sst{A}^p}}
\def\GAn{{G_\sst{A}^n}}
\def\GA{{G_\sst{A}}}
\def\GETEZ{{G_\sst{E}^{\sst{I}=0}}}
\def\GETEO{{G_\sst{E}^{\sst{I}=1}}}
\def\GMTEZ{{G_\sst{M}^{\sst{I}=0}}}
\def\GMTEO{{G_\sst{M}^{\sst{I}=1}}}
\def\lamd{{\lambda_\sst{D}^\sst{V}}}
\def\lamn{{\lambda_n}}
\def\lams{{\lambda_\sst{E}^{(s)}}}
\def\bvz{{\beta_\sst{V}^0}}
\def\bvo{{\beta_\sst{V}^1}}
\def\Gdip{{G_\sst{D}^\sst{V}}}
\def\GdipA{{G_\sst{D}^\sst{A}}}
\def\fks{{F_\sst{K}^{(s)}}}
\def\FIS{{F_i^{(s)}}}
\def\fpi{{F_\pi}}
\def\fk{{F_\sst{K}}}

\def\RAp{{R_\sst{A}^p}}
\def\RAn{{R_\sst{A}^n}}
\def\RVp{{R_\sst{V}^p}}
\def\RVn{{R_\sst{V}^n}}
\def\rva{{R_\sst{V,A}}}
\def\xbb{{x_B}}

\def\delalr{{{delta\alr\over\alr}}}
\def\pbar{{\bar{p}}}
\def\lamchi{{\Lambda_\chi}}

\begin{titlepage}

\hfill {INT \#DOE/ER/40561-317-INT97-00-169}\\
\hphantom{xxx} \hfill {MKPH-T-97-10}

\vspace{1.0cm}

\begin{center}
{\large {\bf Nucleon Vector Strangeness Form Factors:\\
Multi-pion Continuum and the OZI Rule}}

\vspace{1.2cm}

H.-W. Hammer$^{a,b}$  and
M.J. Ramsey-Musolf$^{a,}$\footnote{On leave from the
Department of Physics, University of Connecticut, Storrs, CT
08629\ \ USA}$^,$\footnote{National Science Foundation
Young Investigator}

\vspace{0.8cm}

$^a$ Institute for Nuclear Theory, 
University of Washington, Seattle, WA 98195, USA\\
$^b$ Universit{\"a}t Mainz, Institut f{\"u}r Kernphysik,
J.-J.-Becher Weg 45, D-55099 Mainz, Germany\\[0.4cm]
\end{center}

\vspace{1cm}

\begin{abstract}
We estimate the $3\pi$ continuum contribution to the nucleon strange quark 
vector current form factors, including the effect of
$3\pi\leftrightarrow\rho\pi$ and $3\pi\leftrightarrow\omega$
resonances. We find the
magnitude of this contribution to be comparable to 
that the lightest strange intermediate states. We also study the isoscalar
electromagnetic form factors, and find that the presence of a
$\rho\pi$ resonance in the multi-pion continuum may generate an
appreciable contribution.\\[0.3cm]
\end{abstract}

\vspace{2cm}
\vfill
\end{titlepage}

\section{Introduction}
\label{sec:intro}

The role of the $s\bar{s}$ sea in the low-energy structure of the nucleon
has received considerable attention recently \cite{Mus94}. In particular, 
several experiments -- both in progress as well as in preparation -- 
will probe the strange sea by measuring the nucleon's strange quark vector 
current form factors, $\GES$ and $\GMS$ \cite{McK89}. 
Motivated by these prospective measurements,
a number of theoretical predictions for $\GES$ and $\GMS$ have been 
made. While attempts to carry out a first principles QCD calculation using 
the lattice are still in their infancy \cite{Liu95}, 
the use of effective hadronic
models have provided a more tractable approach to treating the 
non-perturbative physics responsible for the form factors 
\cite{MuI96,MuB94,GeI97,Koe92,Jaf89,HaM96,For96,Mus95,Mei97}. 
Unfortunately,
the connection between a given model and the underlying dynamics of
non-perturbative QCD is usually not transparent. Hence, one finds
a rather broad range of predictions for the strange-quark form factors.

\begin{figure}[h]
\epsfysize=16cm
\rotate[r]{\epsffile{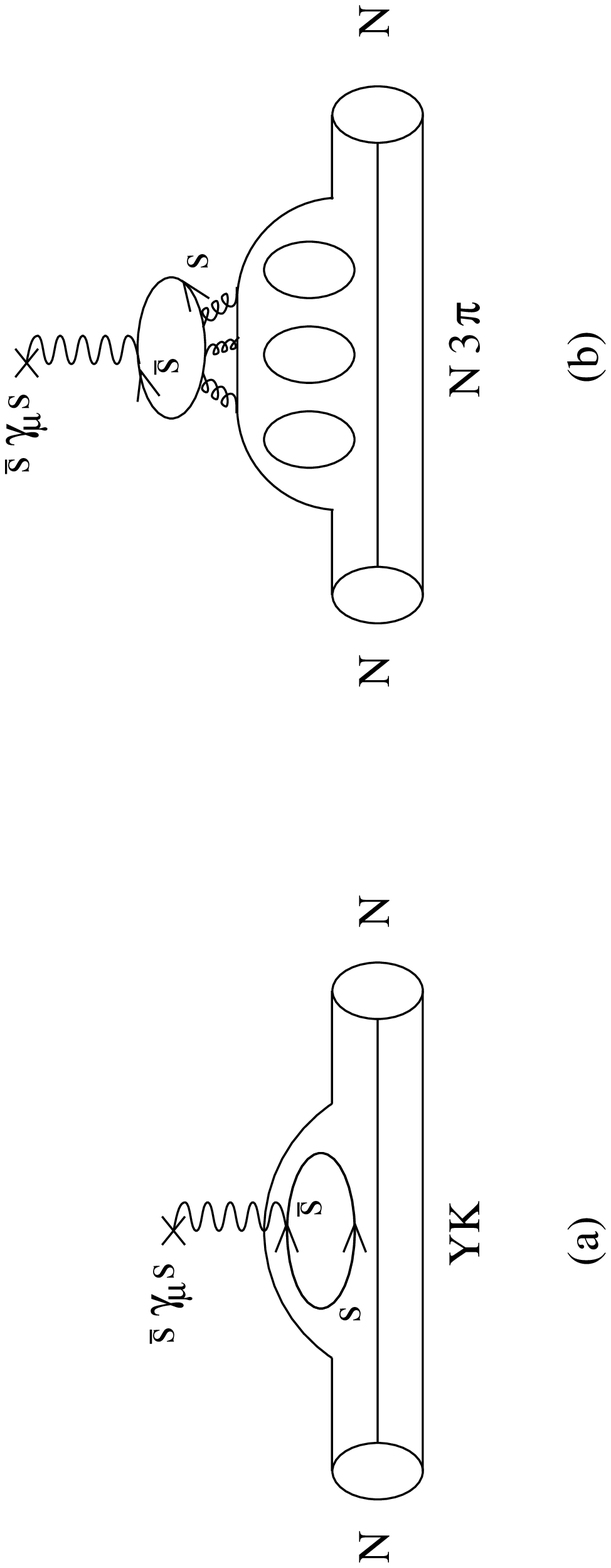}}

\vspace{1.0cm}

\caption{\label{Fig1} Schematic illustration of (a) OZI-allowed
and (b) OZI-violating mesonic contributions
to the strange-quark vector current matrix element in the nucleon.
Curly lines denote gluons (3 is the minimum number required for a vector
current insertion). Un-marked lines and closed loops denote meson and
baryon valence quarks.}
\end{figure}

One may hope, nevertheless, to derive some qualitative insights into
the mechanisms governing $\GES$ and $\GMS$ by using models. Indeed,
one issue which models may address is the validity of using the OZI rule
\cite{Mei97} as a guide to the expected magnitude of the 
strange-quark form factors.
With the OZI rule in mind, most model calculations have assumed that
$\GES$ and $\GMS$ are dominated by OZI-allowed hadronic proceses
in which the nucleon fluctuates into intermediate
states containing valence $s$- and $\bar{s}$-quarks (Fig.~\ref{Fig1}a). 
In a recent quark model calculation, 
Geiger and Isgur \cite{GeI97} have shown -- at second order
in the strong hadronic coupling -- that performing a sum over a tower of
such OZI-allowed states leads to small values for the leading
strangeness moments. While conclusions based on a second-order 
calculation may be questioned \cite{MuH97}, the assumption that 
such OZI-allowed
processes dominate the form factors is largely un-tested\footnote{Pole
model treatments have incorporated the possibility of OZI-violation.
We discuss these analyses in more detail below.}. 

In what follows, we analyze the validity of this assumption by studying
the contribution from the $3\pi$ intermediate state (Fig.~\ref{Fig1}b). The
three pion state is the  lightest state carrying the same quantum numbers 
as the strange quark vector current. As this state also contains no valence 
$s$- or $\bar{s}$-quarks, it represents the simplest case by which to test 
the assumption that the OZI-allowed processes indicated above 
are the most important. On general 
grounds, one might expect the $3\pi$ contribution to be suppressed for two 
reasons: (a) its contribution depends on the OZI-violating
matrix element $\bra{3\pi}\bar{s}\gamma_\mu s\ket{0}$ and (b) chiral 
power counting implies the presence of  additional factors of $p/4\pi F_\pi$ 
(where $p$ is a small momentum or mass) as compared to contributions from 
two particle intermediate states. The latter expectation is supported by
the recent calculation of Ref. \cite{BKM96}, in which the isoscalar
EM form factor spectral functions were analyzed to ${\cal O}(q^7)$ in
heavy baryon chiral perturbation theory (CHPT). In that analysis, the $3\pi$
contribution was found to lie well below the corresponding $2\pi$
continuum contribution to the isovector EM spectral functions. Moreover,
the isoscalar spectral functions computed in Ref. \cite{BKM96} rise 
smoothly from zero at threshold ($t=9\mpis$) and show no evidence of
of a near-threshold singularity enhancement as occurs in the $2\pi$
isovector continuum. 

Given the foregoing arguments, 
one would expect the $3\pi$ continuum contributions to the strangeness
form factors to be negligible. We find, however, that the presence of
$3\pi\leftrightarrow\rho\pi$ and $3\pi\leftrightarrow\omega$
resonances in the amplitudes
$\langle N\bar{N}|3\pi\rangle$ and $\bra{3\pi}\bar{s}\gamma_\mu s\ket{0}$
may enhance the $3\pi$ continuum contribution to a level comparable to
that of typical two-particle, OZI-allowed continuum effects. Our estimate
of the $3\pi\leftrightarrow\omega$ resonance contribution is similar
in spirit to that of previous studies \cite{Jaf89,HaM96,For96}. 
The impact of the
$3\pi\leftrightarrow\rho\pi$ resonance, however, has not been considered
previously. Our estimate of
this contribution depends crucially on the measured partial width for
the decay $\phi\to\rho\pi$, which signals the presence of 
non-negligible OZI-violation in the matrix element 
$\bra{3\pi}\bar{s}\gamma_\mu s\ket{0}$. The inclusion
of a $3\pi\leftrightarrow\rho\pi$ resonance itself is well-founded on
phenomenological grounds, as we discuss below. Consequently, we also
consider its role in the isoscalar EM channel. In this channel, the 
$\rho\pi$ structure
in the $3\pi$ continuum appears to have a particularly significant
impact on the isoscalar magnetic moment, and suggests that conventional
pole analyses of the isoscalar form factors may need modification to
account explicitly for continuum effects\footnote{The recent work of
Ref. \protect\cite{Mei97} reaches a similar conclusion regarding
conventional pole analyses.}.

In arriving at these conclusions, we have relied on a calculation 
requiring model-dependent assumptions. In the absence
of a more tractable first principles (QCD) or effective theory approach,
the use of a model allows us to compare the $3\pi$ contribution to the
lowest lying OZI-allowed contribution computed using the same model 
as Refs. \cite{MuB94,MuH97}. 
In making this comparison, we draw on the framework
of dispersion relations, which affords a systematic means for identifying
various hadronic contributions to the form factors of interest. In
focusing on the $3\pi$ contribution, we discuss the phenomenological
justification for including the $\omega$ and
$\rho\pi$ resonances, and present our model calculation for 
these contributions. We conclude with a
discussion of the results and their implications.

\section{Dispersion Relations}
\label{sec:disp}

The framework of dispersion relations is well suited for the study of
multi-meson contributions to the nucleon form factors. Following Refs. 
\cite{FGT58,MuH97}, we work in the $N\bar{N}$ production channel, 
where these form factors are defined as
\begin{equation}
\bra{N(p);\bar{N}(\pbar)} J_\mu^{(a)}\ket{0}={\bar U}(p)
\left[F_1^{(a)}(t)\gamma_\mu+{iF_2^{(a)}(t)\over 2\mn}\sigma_{\mu\nu}
P^\nu \right]V(p)\ \ \ ,
\label{eq:ffa}
\end{equation}
where $P^\mu=(p+\pbar)^\mu$, $t=P^2$, $U$ ($V$) is a nucleon
(anti-nucleon) spinor, and \lq\lq $(a)$" denotes the flavor channel
[$J_\mu^\sst{EM}(I=0)$ or $\bar{s}\gamma_\mu s$].
The Dirac and Pauli strangeness form factors
are related to $G_\sst{E}^{(a)}$ and $G_\sst{M}^{(a)}$ as 
\begin{eqnarray}
\label{eq:geff}
G_\sst{E}^{(a)}&=&F_1^{(a)}-\tau F_2^{(a)}\ \ \ ,\\
\label{eq:gmff}
G_\sst{M}^{(a)}&=&F_1^{(a)}+F_2^{(a)}\ \ \ ,
\end{eqnarray}
where $\tau=-t/4\mns$. For elastic processes involving the nucleon, one
has $t=(p-p')^2$, where $p$ and $p'$ are the initial and final nucleon
momenta. 

Since the value of $F_1^{(a)}$ at $t=0$ is rigorously known -- it is
just the nucleon's net isoscalar EM charge ($=1/2$) or net strangeness
($=0$) --  we use a subtracted
dispersion relation for $F_1^{(a)}$. Since we want to determine
the value of $F_2^{(a)}$ at $t=0$, we use an un-subtracted dispersion relation
for the latter\footnote{The use of additional subtractions in the dispersion
relation for $F_1^{(a)}$ would require knowledge of the second and higher
moments of the form factor. Since we wish to study the mean square Dirac 
strangeness radius, we restrict ourselves to a single subtraction in this 
case. For a discussion of theoretical considerations regarding
subtractions and convergence, see Ref. \protect\cite{MuH97} and
references therein.}. Hence, we write
\begin{eqnarray}
\label{eq:dispa}
F_1^{(a)}(t)&=&{t\over\pi}\int_{t_0}^\infty\ dt' {\hbox{Im}\ F_1^{(a)}(t')
             \over t'(t'-t-i\epsilon)}\ \ \ ,\\
\label{eq:dispb}
F_2^{(a)}(t)&=&{1\over\pi}\int_{t_0}^\infty\ dt' {\hbox{Im} F_2^{(a)}(t')
             \over t'-t-i\epsilon}\ \ \ ,
\end{eqnarray}
where $t_0$ begins a cut along the real $t$-axis associated with the
threshold for a given physical state. In what follows, we will be 
particularly interested in the mean square radius and anomalous 
magnetic moment associated with the currents $J_\mu^{(a)}$:
\begin{eqnarray}
\label{eq:rad}
\rho^{(a)} & = & \left.{d F_1^{(a)}\over d\tau}\right|_{t=0} = 
-{2\over 3}\mns \langle r^2\rangle_{(a)} \\
\label{eq:mm}
\kappa^{(a)} & = & F_2^{(a)}(t=0) \ \ \ .
\end{eqnarray}
The corresponding dispersion integrals for the dimensionless mean square
radius and magnetic moment are
\begin{eqnarray}
\label{eq:dispr}
\rho^{(a)} & = & -{4\mns\over\pi} \int_{t_0}^\infty \ dt
        {\hbox{Im}\ F_1^{(a)}(t)\over t^2}\ \ \ ,\\
\label{eq:dispk}
\kappa^{(a)} & = & {1\over\pi}\int_{t_0}^\infty\ dt {\hbox{Im} F_2^{(a)}(t)
        \over t}\ \ \ .
\end{eqnarray}

The spectral decomposition of the spectral functions $\hbox{Im}\ 
F_i^{(a)}(t)$ is readily obtained by applying the LSZ formalism to the 
absorptive part of the matrix element in Eq. (\ref{eq:ffa}) \cite{FGT58}:
\begin{eqnarray}
\label{eq:lsz}
&&\hbox{Im} \bra{N(p);\bar{N}(\pbar)} J_\mu^{(a)}\ket{0}\quad\rightarrow\\
&&\qquad\qquad
{\pi\over\sqrt{Z}} (2\pi)^{3/2}{\cal N}\sum_n\bra{N(p)}\bar{J}_N(0)
\ket{n}\bra{n}J_\mu^{(a)}\ket{0} V(\pbar)\delta^4(p+\pbar-p_n)\ \ \ ,
\nonumber
\end{eqnarray}
where ${\cal N}$ is a nucleon spinor normalization factor, $Z$ is the 
nucleon's wavefunction renormalization constant, and $J_N$ is
a nucleon source.

\begin{figure}[htb]
\epsfxsize=13cm
\epsffile{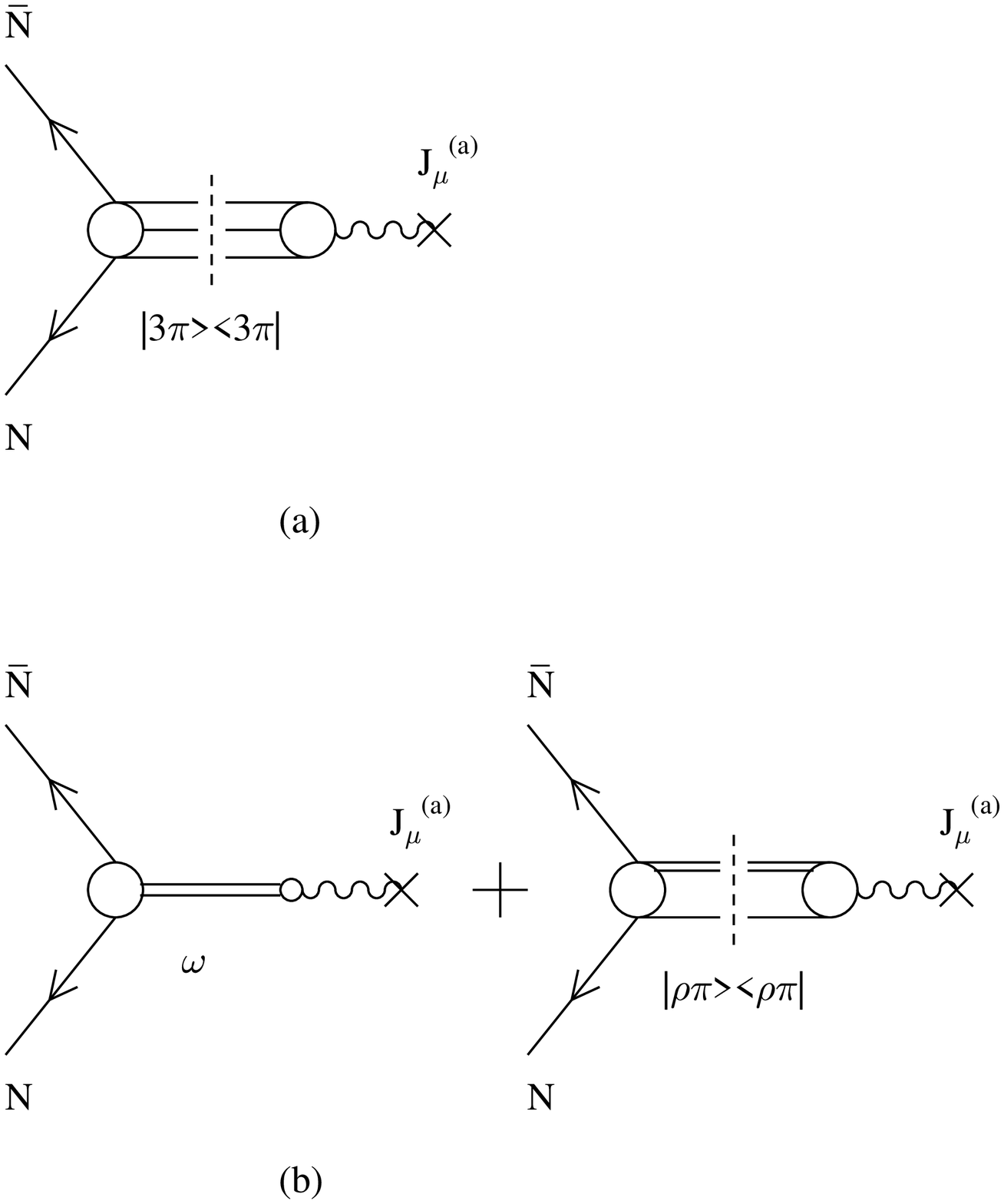}

\vspace{-1cm}

\caption{\label{Fig2} Schematic illustration of (a) the $3\pi$ contribution
to the spectral function, as given in Eq. (\ref{eq:lsz}), and (b) the
model approximation given by Eq. (\ref{eq:sfm}).
Single solid lines denote pions
and double line in (b) denotes a vector meson. Right hand parts of diagrams
represent matrix element to produce a $I^G(J^{PC})=0^-(1^{--})$ state
$\ket{n}$ ($=\ket{3\pi},\ \ket{\omega},\
\mbox{ or }\ket{\rho\pi}$) from the vacuum through
the isoscalar EM or strangeness current. Left hand side denotes
$n \to N\bar{N}$ scattering amplitude.}
\end{figure}

The states $\ket{n}$ of momentum $p_n$ are stable with respect to the
strong interaction ({\em i.e.}, no vector meson resonances) and carry the
same quantum numbers as $J_\mu^\sst{EM}$ and $\bar{s}\gamma_\mu s$:
$I^G(J^{PC})=0^{-}(1^{--})$.
The lightest such state is the three pion state, with a physical threshold
of $t_0=9\mpis$. The contribution from this state to the spectral function
via the decomposition of Eq. (\ref{eq:lsz})
is illustrated diagramatically in Fig. \ref{Fig2}a.
In order of successive thresholds, the next allowed purely mesonic states
are the $5\pi$, $7\pi$, $K\bar{K}$, $9\pi$, $K\bar{K}\pi$, 
$\ldots$. In the baryonic
sector, one has $N\bar{N}$, $\Lambda\bar{\Lambda}$, $\ldots$. One may also
consider states containing both mesons and baryons. 

Most existing hadronic calculations of the strangeness form factors
have either (a) included only the $K\bar{K}$ and $Y
\bar{Y}$ states in the guise of loops -- even though they are not
the lightest such allowed states -- or (b) approximated the entire sum of
Eq. (\ref{eq:lsz})
by a series of poles. In the case of loops, it was shown in Ref.
\cite{MuH97} that $YK$ loop calculations -- which treat the $K\bar{K}$ and 
$Y\bar{Y}$ contributions
together -- are equivalent to the use of a dispersion relation in which
(i) the scattering amplitude $\bra{N(p)}\bar{J}_N(0)\ket{n}V(\pbar)$
is computed in the Born approximation and (ii) the matrix element
$\bra{n}J_\mu^{(a)}\ket{0}$ is taken to be point like, that is,
the $t$-dependence of the associated form factor is neglected. As discussed
in Refs. \cite{MuH97,FGT58}, 
both approximations are rather drastic. Indeed, considerations
of unitarity imply the presence of important meson rescattering corrections
(higher-order loop effects) in the scattering amplitude. Moreover, the
amplitude may also contain vector meson resonances, such as 
$K\bar{K}\leftrightarrow\phi$. In the case of the 
$K\bar{K}$ contribution to the leading strangeness moments, 
it was also shown that the inclusion of a realistic kaon
strangeness form factor can significantly affect the results \cite{MuH97}.

Our aim at present is to compare the contributions arising from the
states $\ket{K\bar{K}}$ and $\ket{3\pi}$ as they enter the decomposition
of Eq. (\ref{eq:lsz}).
With regard to the $3\pi$ contribution, a careful study of $N\bar{N}\to
3\pi$ or $N\pi\to N\pi\pi$ scattering data would be needed to give a
realistic determination of both rescattering and resonance contributions.
The resonances include $3\pi\leftrightarrow V$ 
and $3\pi\leftrightarrow V'\pi$, where $V$ is a $0^{-}(1^{--})$ vector
meson such as the $\omega$ or $\phi$, and where $V'$ is $1^{+}(1^{--})$
vector meson such as the $\rho$. 
Such a study would involve fitting the amplitudes in the physical region
and performing an analytic continuation to the un-physical region appropriate
to the dispersion relation. The feasibility of carrying out this analysis,
given the current state of multipion-nucleon scattering data, is unclear.
In addition, the use of Eq. (\ref{eq:lsz}) requires the matrix element
$\bra{3\pi} J_\mu^{(a)}\ket {0}$, for which -- in the strangeness
channel -- no data currently exists. 
In order to make any statement about the scale of the 
$3\pi$ contribution, then, it appears that one must introduce model
assumptions.

\section{Model Calculation}
\label{sec:model}

In the present context, we seek to determine whether any structure
exists in the $3\pi$ continuum which might enhance its contribution
over the scale one expects based on chiral symmetry \cite{BKM96}
and the OZI-rule. Pure resonance processes ($3\pi\leftrightarrow V$,
{\em etc.}) have been considered previously in the pole model analyses
of Refs. \cite{MuI96,Jaf89,HaM96,For96,Mus95}. 
In these models, the form factors are approximated by
a sum over poles associated with the lightest $0^{-}(1^{--})$ vector mesons;
no explicit mention is made of the intermediate state scattering amplitudes
or form factors.
When applied to the isoscalar EM form factors, this approximation yields
a value for the $\phi$-nucleon coupling significantly larger than 
suggested by the OZI
rule: $g_{\phi NN}/g_{\omega NN}\approx -1/2$. The corresponding
predictions for the strangeness form factors are surprisingly large, 
especially in the case of the strangeness radius \cite{comm}.

In the present study, we seek to determine the existence
of important OZI-violating contributions
to the $F_i^{(s)}(t)$ without relying exclusively on the pole approximation
and the associated large $\phi NN$ coupling. Instead, we adopt the following
model approximation for the $3\pi$ contribution to the spectral function:
\begin{equation}
\hbox{Im}\ F_i^{(a)}(t)^{3\pi}=a_i^{(a)}\delta(t-t_\omega)+
\hbox{Im}\ F_i^{(a)}(t)^{\rho\pi}\ \ \ ,
\label{eq:sfm}
\end{equation}
where $t_\omega=m_\omega^2$. The first term constitutes a model for the
$3\pi\leftrightarrow\omega$ contribution in which a narrow resonance
approximation is made for the $3\pi$ spectral function in the vicinity
of $t_\omega$: $\langle N\bar{N}\vert 3\pi\rangle\bra{3\pi}J_\mu^{(a)}\ket{0}
\to\langle N\bar{N}\vert\omega\rangle\bra{\omega}J_\mu^{(a)}\ket{0}$.
The second term models the $3\pi\leftrightarrow\rho\pi$ contribution. 
The latter, which becomes non-zero only for $t\geq (m_\rho+m_\pi)^2$,
assumes that the $3\pi$ continuum can be approximated by a $\rho\pi$ 
state in this kinematic regime: $\langle N\bar{N}\vert
3\pi\rangle\bra{3\pi}J_\mu^{(a)}\ket{0}\to\langle 
N\bar{N}\vert\rho\pi\rangle\bra{\rho\pi}J_\mu^{(a)}\ket{0}$.
The content of this model for $\hbox{Im}\ F_i^{(a)}(t)^{3\pi}$ is 
illustrated diagramatically in Fig. \ref{Fig2}b.

The rationale for our model rests on several observations.
We first note that of the states appearing in the spectral
decomposition of Eq. (\ref{eq:lsz}), only the $\ket{3\pi}$ and $\ket{5\pi}$
can resonate entirely into an $\omega$, since the cuts for the
other states occur for $t_0> m_\omega^2$ and since $\Gamma_\omega/
m_\omega \ll 1$. Experimentally, it is well known that one finds
a strong peak in $e^+e^-\to 3\pi$ near $\sqrt{s}=m_\omega$ \cite{Dol91}. 
On the other
hand, there does not exist -- to our knowledge -- convincing evidence
for a strong coupling of the $\omega$ to five pions. Moreover, 
$\pi^+\pi^-\pi^0$ invariant mass distributions for $e^+e^-\to
5\pi$ events display a peak near $m_\omega$ \cite{Cor81}, 
indicating that the presence
of the $\omega$ in the $5\pi$ channel arises primarily through the
$3\pi\leftrightarrow\omega$ resonance (for $t>1\mbox{ GeV}^2$). 
Hence, one has reason to assume 
that the presence of a pure $\omega$ resonance in $\hbox{Im}\ F_i^{(a)}(t)$ 
arises dominantly through the $3\pi$ intermediate state. Given
the relatively narrow width of the $\omega$, the narrow resonance
approximation appearing in Eq. (\ref{eq:sfm}) also appears to be justified.

Several experimental facts motivate the inclusion of the $\rho\pi$ term 
in our model. First, there exists a strong peak in $e^+e^-\to 3\pi$
near $\sqrt{s}=m_\phi$ and a large $\rho\pi$ continuum for $\sqrt{s}
> 1.03$ GeV \cite{Dol91}. 
Second, the $\phi$ has a non-negligible partial width 
$\Gamma(\phi\to\rho\pi)$ whereas the partial width for decay into
an un-correlated $3\pi$ state is a more than a factor of four smaller
\cite{PDG}. Together, these observations suggest that, for $t\simge 1 
\mbox{ GeV}^2$, the matrix element $\bra{3\pi} J_\mu^{(a)}\ket{0}$
is dominated by processes in which the $3\pi$ resonate to a $\rho\pi$
and then -- for $t\approx m_\phi^2$ -- to the $\phi$. Since the $\phi$
is almost a pure $s\bar{s}$ state, one would expect the $3\pi\leftrightarrow
\rho\pi\leftrightarrow\phi$ mechanism to generate a non-negligible
contribution to $\bra{3\pi}\bar{s}\gamma_\mu s\ket{0}$. 

Regarding the $N\bar{N}\to 3\pi$ scattering amplitude, 
one observes a pronounced $\rho\pi$ resonance in the
$3\pi$ channel for $p\bar{p}$ annihilation at rest. The latter has been
seen in bubble chamber data \cite{Fos68} and at LEAR \cite{May89}
(for a review, see Ref. \cite{AmM91}).
From analyses of these data, one finds the $\rho\pi$ branching
ratio to be roughly three times that for the direct (non-resonant)
$3\pi$ final state. In addition, plots of $3\pi$ yield {\em vs.}
$\pi^+\pi^{-}$ invariant mass display peaks in the vicinity of the
$f_2(1270)$ and $f_2(1565)$ vector mesons as well as the $\rho^0(770)$.
While these data exist only at the $p\bar{p}$ threshold ($t=4\mns$),
they suggest the presence of significant $V'\pi$ resonance structure
elsewhere along the $3\pi$ cut. In our model, then, we consider only
the lightest such resonance ($\rho\pi$) and approximate the state
$\ket{3\pi}$ as the state $\ket{\rho\pi}$ for $t\geq (m_\rho+m_\pi)^2$.
While this approximation entails omission of higher-lying
$V'\pi$ structure,
it is nevertheless sufficient for estimating the possible scale of the
$3\pi$ contribution to the strangeness moments.
We also note in passing that our approximation here is
similar to the one employed in Ref. \cite{GeI97}, where OZI-allowed
multi-pion states, ({\em e.g.}, $K\bar{K}\pi$) were approximated as
two particle states ({\em viz}, $K^{\ast}\bar{K}$).

To calculate the $a_i^{(a)}$ and $\hbox{Im}\, F_i^{(a)}$ we require
the strong meson-nucleon vertices. For the $NN\pi$ coupling,
we employ the linear $\sigma$ model, and for the $NNV$ interaction,
we use the conventional vector and tensor couplings (see, {\em e.g.}
Ref. \cite{Hol89}). The corresponding Lagrangians are
\begin{eqnarray}
\label{eq:lpinn}
{\cal L}_{N N\pi} &=& -i\gpnn \bar{N} {\vec\tau}\cdot{\vec\pi}\gamma_5 N \\
\label{eq:lrhonn}
{\cal L}_{N N V} &=& -\bar{N}\left[ \gvnn\gamma_\mu - {f_\sst{NNV}
\over 2\mn}\sigma_{\mu\nu}{\partial}^\nu\right] V^\mu N
\ \ \ ,
\end{eqnarray}
where $V^\mu ={\vec\tau}\cdot{\vec\rho}^\mu$ and $V^\mu=\omega^\mu$ 
for the $\rho$ and $\omega$, respectively, and where the derivative 
in Eq. (\ref{eq:lrhonn}) acts only on the $V^\mu$ field. 
 
The pole approximation for the $\omega$-resonance contribution yields
\begin{eqnarray}
\label{eq:omega}
a_1^{(a)} &=& \gonn/f_\omega^{(a)} \\
a_1^{(a)} &=& \gonn/f_\omega^{(a)} 
\end{eqnarray}
where the $f_V^{(a)}$ are defined via
\begin{equation}
\bra{0} J_\mu^{(a)}\ket{V(q,\varepsilon)} =  {M_V^2\over f_V^{(a)}}
\varepsilon_\mu
\ \ \ .
\end{equation}
A non-zero value for $f_\omega^{(s)}$ arises from the small $s\bar{s}$
component of the $\omega$; it may be
determined using the arguments of Ref. \cite{Jaf89}
as discussed below. For the strong couplings, we employ values taken from
fits to $NN$ scattering \cite{Hol89}: 
$\gonn=15.853$ and $\fonn=0$. The resultant
$\omega$ contributions to the strangeness radius and magnetic moment do not
differ appreciably if we use, instead, the strong couplings determined from
pole analyses of isoscalar EM form factors \cite{Mer96}. 
We note that this method for
including the $\omega$ contribution is identical to that followed in
Refs. \cite{Jaf89,HaM96,For96,MuI96}.

\begin{figure}[htb]
\epsfclipon
\epsfxsize=14cm
\epsffile{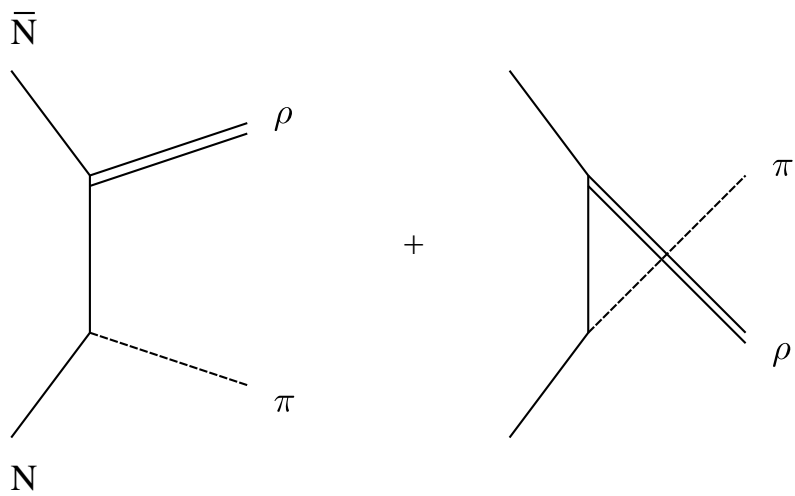}

\vspace{-1.0cm}

\caption{\label{Fig3} Born diagrams for the $\rho\pi\to N\bar{N}$ 
amplitude.}
\end{figure}

When evaluating the function $\hbox{Im}\ F_i^{(a)}(t)^{\rho\pi}$, 
we follow the philosophy of Ref. \cite{GeI97} and evaluate the
$N\bar{N}\to\rho\pi$ amplitude in the Born approximation (Fig. \ref{Fig3}), 
neglecting possible meson-meson correlations \cite{Mul95}. 
In addition, we require 
knowledge of the matrix element $\bra{\rho\pi}J_\mu^{(a)}\ket{0}$, 
which one may parameterize as
\begin{equation}
\label{eq:jrhopi}
\bra{\rho^b(\varepsilon, k)\ \pi^c(q)} J_\mu^{(a)}\ket{0}=
{g_{\rho\pi}^{(a)}\over\mro}\delta^{bc}\epsilon_{\mu\alpha\beta\lambda}
q^\alpha k^\beta \varepsilon^{\lambda\ast}\ \ \ .
\end{equation}
In the case of the isoscalar EM current, one may obtain the value of
$g_{\rho\pi}^{I=0}(t=0)$ from the radiative decay of the $\rho$.
For purposes of computing the dispersion integrals, one also
requires the behavior of $g_{\rho\pi}^{I=0}(t)$ away from the photon
point. Since there exist no data for $t\not= 0$, one must rely on
a model for this kinematic region. A similar statement
applies in the strangeness channel, since there exist no data for
$g_{\rho\pi}^{(s)}(t)$ for any value of $t$. We therefore follow
Ref. \cite{GoM96} and employ a vector meson dominance (VMD) model for
the transition form factors. The VMD model yields
\begin{equation}
\label{eq:vdmff}
{g_{\rho\pi}^{(a)}(t)\over m_\rho} = 
\sum_{V=\omega,\phi}{(G_{\sst{V}\rho\pi}/f_V^{(a)})
\over 1-t/M_V^2-i\Gamma_V t/ M_V^3}\ \ \ ,
\end{equation}
where the $G_{\sst{V}\rho\pi}$ 
are strong $V\rho\pi$ couplings, the $1/f_V^{(a)}$
give the strength of the matrix element $\bra{V} J_\mu^{(a)}\ket{0}$,
and $\Gamma_V t/ M_V^2$ gives an energy-dependent width for vector 
meson $V$ as used in Ref. \cite{FeS81}. The $G_{\sst{V}\rho\pi}$
have been determined in Ref. \cite{GoM96} using the measured partial width
$\Gamma(\phi\to\rho\pi)$ and a fit to radiative decays of vector mesons in
conjunction with the VMD hypothesis. The isoscalar EM constants
$f_{\omega,\phi}^{I=0}$ are well known. Using the arguments of Ref.
\cite{Jaf89}, we may determine the corresponding values in the strangeness
sector: $1/f_\omega^{(s)}\approx -0.2/f_\omega^{I=0}$ and 
$1/f_\phi^{(s)}\approx -3/f_\phi^{I=0}$. The value for $1/f_\phi^{(s)}$
is essentially what one would expect were the $\phi$ to be a pure
$s\bar{s}$ state \cite{GoM96}, while the small, but non-zero, value of 
$1/f_\omega^{(s)}$ accounts for the small $s\bar{s}$ component of 
the $\omega$. The resulting values for the $g_{\rho\pi}^{(a)}(0)$ are
0.53 and 0.076 in the EM and strangeness channels, respectively.

Using Eqs. (\ref{eq:lpinn}-\ref{eq:vdmff}) as inputs, we obtain the 
spectral functions $\hbox{Im}\,F_i^{(a)}(t)^{\rho\pi}$ for $t\geq4\mns$, 
following the procedures 
outlined in the Appendix of Ref. \cite{MuH97}:
\begin{eqnarray}
\label{eq:imf1}
\hbox{Im}\ F_1^{(a)}(t)^{\rho\pi}&=&\hbox{Re}\,
\frac{g_{\rho\pi}^{(a)}(t)^*}{m_\rho}
\frac{\gpnn Q^2 }{16\pi\sqrt{t} P}\left[\grnn\frac{\mn t}{2 P^2} Q_2(z)
+\frac{f_\sst{\rho NN}}{2\mn}\left\{ \frac{\mns\, t}{P^2}Q_2(z)
\vphantom{\left(1+\frac{P^2}{\mns}\right)}\right.\right. \\ & &\left.\left.
+\frac{t}{3}[2 Q_0(z)+Q_2(z)] -\frac{\mns}{P Q}\left(2 K^2-\frac{t}{2}
\right)\left(1+\frac{P^2}{\mns}\right)Q_1(z) \right\}\right]\ \ , 
\nonumber \\ \label{eq:imf2}
\hbox{Im}\ F_2^{(a)}(t)^{\rho\pi}&=&\hbox{Re}\,
\frac{g_{\rho\pi}^{(a)}(t)^*}{m_\rho}
\frac{\gpnn Q^2 }{16\pi\sqrt{t} P}\left[\grnn\mn\left\{\frac{2}{3}
[Q_2(z)+2 Q_0(z)]
\vphantom{\frac{\mns}{P^2}}\right.\right. \\ & &\left.\left. 
\vphantom{\frac{\mns}{P^2}} -\frac{t}{2 P^2} Q_2(z)\right\}
-\frac{f_\sst{\rho NN}}{2\mn}\left\{ \frac{\mns\, t}{P^2}
Q_2(z) -\frac{\mns}{P Q}\left(2 K^2-\frac{t}{2}\right) Q_1(z) \right\}
\right] \ \ , \nonumber
\end{eqnarray}
where 
\begin{eqnarray}
z &=& \frac{m_\pi^2+ m_\rho^2 -2 Q^2-t-2\sqrt{(Q^2
+m_\pi^2)(Q^2+ m_\rho^2)}}{8 P Q} \ \ , \\
P&=&\sqrt{t/4-\mns}\ \ , \\
Q&=&\frac{1}{2\sqrt{t}}\sqrt{t^2+\left(m_\rho^2- m_\pi^2\right)^2
-2t\left(m_\pi^2+ m_\rho^2\right)}\ \ , \\
K^2&=&\frac{1}{4}\left(\sqrt{Q^2+ m_\rho^2}-\sqrt{Q^2+ m_\pi^2}\right)^2
-Q^2\ \ ,
\end{eqnarray}
and the $Q_i(z)$, $i=0,1,2$ are the Legendre functions of the second kind.
Since we have made the two-particle $\rho\pi$ approximation for the $3\pi$
continuum, the spectral functions become non-zero only for 
$t>t_0=(\mro+\mpi)^2$, rather than $t_0=9\mpi^2$. Moreover, because $t_0$
lies below the physical $N\bar{N}$ production threshold, the imaginary
parts from Eqs. (\ref{eq:imf1}) 
and (\ref{eq:imf2}) have to be analytically continued into the
unphysical region $(\mro+\mpi)^2 \leq t \leq 4\mns$. 
In order to evaluate the dispersion integrals using these spectral functions, 
we must make an additional assumption regarding the relative phases of the 
quantities entering Eqs. (\ref{eq:imf1}, \ref{eq:imf2}). 
In the absence of sufficient experimental data for 
the $N\bar{N}\to\rho\pi$ amplitude and for the $g_{\rho\pi}^{(a)}(t)$,
we have no unambiguous
way of determining the relative phases of these two quantities.
We therefore follow Ref. \cite{MuH97} and replace the scattering amplitude
and form factors in Eqs. (\ref{eq:imf1}, \ref{eq:imf2})
by their magnitudes -- a procedure which
yields an upper bound on the magnitude of the spectral functions. We also
take the overall sign for the spectral functions from our model,
although this choice has no rigorous justification. 
Explicit numerical results
obtained under these assumptions, together with the VDM hypothesis of
Eq. (\ref{eq:vdmff}), are given in the following section.

\section{Results and Discussion}
\label{sec:results}

\begin{figure}[htb]
\epsfysize=17cm
\begin{center}
\epsffile{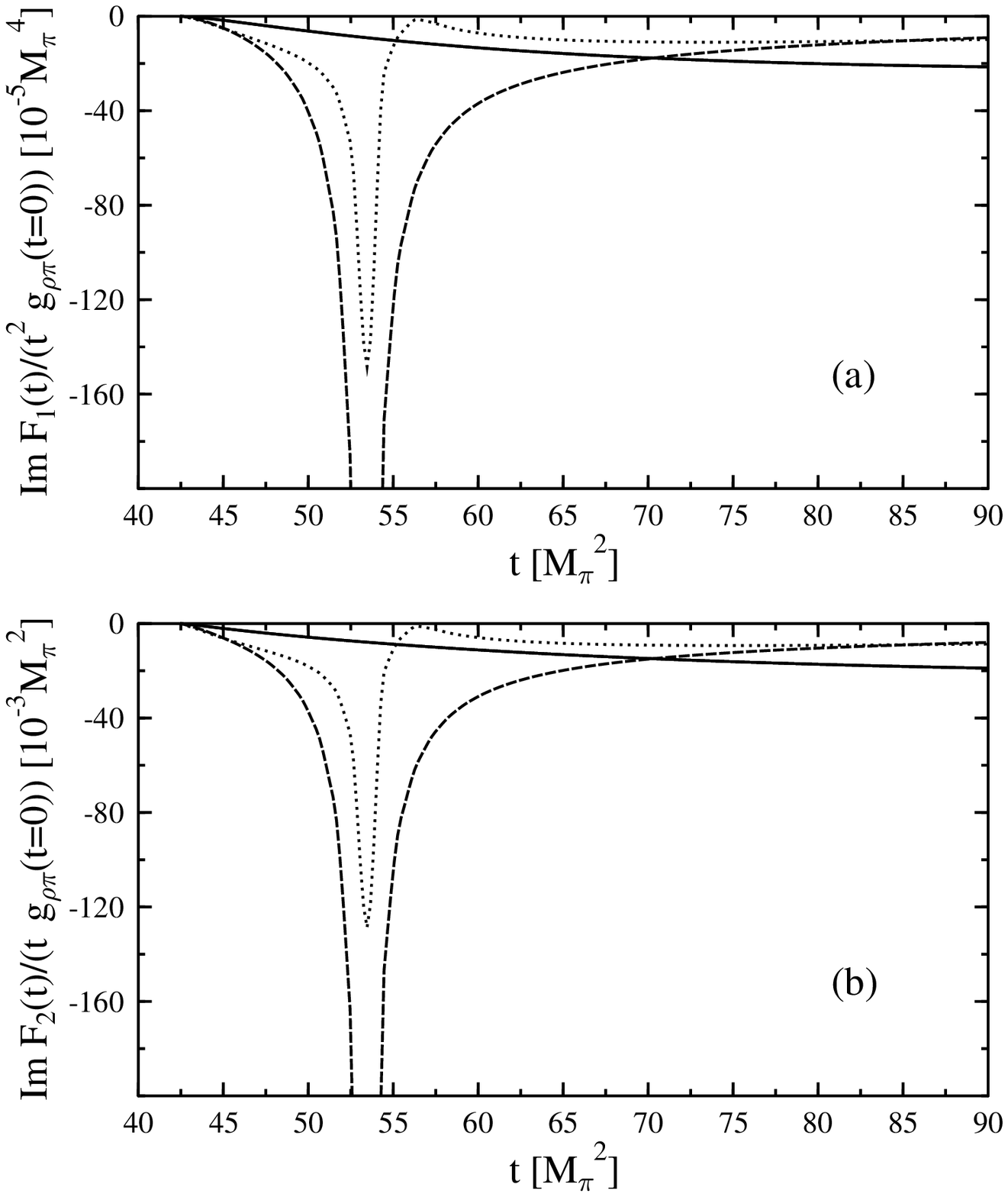}
\end{center}
\caption{\label{Fig4} Weighted spectral functions entering the dispersion
integrals for the mean square radius (a) and anomalous
magnetic moment (b), 
scaled to the value of $g_{\rho\pi}^{(a)}(t=0)$. In the
case of scenario (1) (solid line), the curves for the isoscalar EM and
strangeness channels are identical. For scenario (2), the isoscalar EM
(dotted line) and strangeness (dashed line) spectral functions differ.
The weighted, un-scaled spectral functions are obtained from the curves
by multiplying by $g_{\rho\pi}^{I=0}(0) = 0.53$ and 
$g_{\rho\pi}^{(s)}(0) = 0.076$ as appropriate.}
\end{figure}

In Fig. \ref{Fig4} we plot the spectral functions 
of Eqs. (\ref{eq:imf1}, \ref{eq:imf2}), scaled to the
value of $g_{\rho\pi}^{(a)}(t=0)$ and weighted by powers of $1/t$
as they enter the radius and magnetic moment [Eqs. (\ref{eq:dispr},
\ref{eq:dispk})]. We do not display the region containing the
contribution from the first term in Eq. (\ref{eq:sfm}) as it simply
yields a $\delta$-function at $t\approx 25\mpis$.
We consider two scenarios for
illustrative purposes: (1) a point-like form factor at the $\rho\pi$
current vertex, {\em i.e.}, $g_{\rho\pi}^{(a)}(t)=g_{\rho\pi}^{(a)}(0)$
and (2) $g_{\rho\pi}^{(a)}(t)$ given by Eq. (\ref{eq:vdmff}). 
For scenario (1), the plots for both the isoscalar EM and 
strangeness channels are identical. In this case, the spectral functions 
rise smoothly from zero at the $\rho\pi$ threshold
[$t_0=(\mro+\mpi)^2$] and exhibit no structure suggesting any enhancement
over the un-correlated $3\pi$ continuum. The structure of the spectral
functions in scenario (2), where we include more realistic $\rho\pi$ form
factors, is markedly different. In both the EM and strangeness channels,
the spectral functions contain a strong peak in the vicinity of the
$\phi$ resonance, followed by a subsequent suppression for larger $t$
compared to scenario (1). Although the $\rho\pi$ form factors
contain both $\omega$- and $\phi$-pole contributions, the effect of the
$\omega$ is rather mild since $m_\omega$ lies below the $\rho\pi$
threshold. The $\phi$ peak itself is stronger in the strangeness
as compared to the EM channel, since $1/f_\phi^{(s)}/\approx -3/
f_\phi^{I=0}$.

Using these spectral functions, we compute the contributions to the EM 
isoscalar and strangeness radii and magnetic moments. The results are
given in Table I. For the illustrative purposes, we have listed the
$3\pi\leftrightarrow\omega$ and $3\pi\leftrightarrow\rho\pi$ contributions
to the strangeness form factors
separately. In the case of scenario (1), the dispersion integrals
diverge. Consequently, we only quote results for scenario (2), in
which the VMD $\rho\pi$ form factors render the integrals finite.
We also give the contribution from the lightest intermediate
state containing valence $s$- and $\bar{s}$-quarks ($K\bar{K}$), 
computed using similar assumptions as in the case
of the $\rho\pi$ contribution: (a) the Born approximation for the
$N\bar{N}\to K\bar{K}$ amplitude; (b) the linear SU(3) $\sigma$-model
for the hadronic couplings; and (c) a VMD form factor at the 
$K\bar{K}$-current vertex. We reiterate that the phases of the
$\rho\pi$ contributions in Table I are not certain. 

$$\hbox{\vbox{\offinterlineskip
\def\strut{\hbox{\vrule height 15pt depth 10pt width 0pt}}
\hrule
\halign{
\strut\vrule#\tabskip 0.2cm&
\hfil$#$\hfil&
\vrule#&
\hfil$#$\hfil&
\vrule#&
\hfil$#$\hfil&
\vrule#&
\hfil$#$\hfil&
\vrule#\tabskip 0.0in\cr
& \multispan7{\hfil\bf TABLE I\hfil} & \cr\noalign{\hrule}
& \hbox{Flavor Channel } && \hbox{Source} && \rho && \kappa
& \cr\noalign{\hrule}
& \hbox{I=0, EM} && \hbox{Exp't} && -4.56 && -0.06 & \cr
& \hbox{I=0, EM} && 3\pi\leftrightarrow\rho\pi
        && 0.60 && -0.38 &\cr
& \hbox{Strange} && 3\pi\leftrightarrow\rho\pi
        && 0.86 && -0.44 & \cr
& \hbox{Strange} && 3\pi\leftrightarrow\omega
        && 1.08  && 0 &\cr
&                &&  &&  (1.41)  && (0.04) &\cr
& \hbox{Strange} && K\bar{K} && 0.53 && -0.16 &\cr
\noalign{\hrule}}}}$$
{\noindent\narrower {\bf Table I.} \quad {\small Isoscalar EM and strangeness
dimensionless mean square Dirac radius and anomalous magnetic moment. 
First row gives experimental values for isoscalar EM moments. Second and
third rows give contributions to the EM and strangeness moments
from the $\rho\pi$ intermediate state, computed using the Born 
approximation for the $N\bar{N}\to\rho\pi$ amplitude and a vector meson 
dominance model $\rho\pi$ form factor. Fourth and fifth rows list
$\omega$-resonance contribution to strangeness moments. Numbers in
parentheses correspond to strong couplings obtained from fits
to isoscalar EM form factors.
Final row gives lightest OZI-allowed 
intermediate state contribution to the strangeness moments, computed
using the same approximations as for the $\rho\pi$ case. To convert
$\rho$ to $\langle r^2\rangle$, multiply $\rho$ by -0.066 fm$^{-2}$.}
\smallskip}

The results for the strangeness radius and magnetic moment indicate
that the magnitude of the $3\pi$ contribution, arising 
via the $\omega$ and  $\rho\pi$ resonances,
is similar to that of the $K\bar{K}$ contribution. At least within the
present framework, then, we find no reason to neglect the lightest
OZI-violating intermediate state contribution, 
as is done in nearly all existing hadronic model
calculations. Moreover, the magnitudes of the $\omega$ and $\rho\pi$ 
contributions are comparable for the radius, while the $\rho\pi$
term gives a significantly larger contribution to $\kappa^s$.
This result appears to depend rather crucially on the
presence of the $\phi$-meson pole in $g_{\rho\pi}^{(s)}(t)$ which
enhances the spectral functions in the region where they are weighted
most heavily in the dispersion integral.

We also find that while the $\rho\pi$ contribution to the
isoscalar EM charge radius is small compared to experimental value,
the isoscalar EM magnetic moment is not. This results suggests that
a re-analysis of the $F_i^{I=0}(t)$ ought to be performed, including
not only the effects of sharp $0^-(1^{--})$ resonances but also those
from the $\rho\pi$.

It is instructive to compare our results with those obtained for the
isoscalar EM spectral functions using CHPT and those obtained in the
pure pole approximation. A direct comparison with the former is difficult,
since the chiral expansion is valid only for $\sqrt{t}\ll 4\pi F_\pi$,
whereas the $\rho\pi$ contribution to the spectral function becomes
non-zero only for $\sqrt{t}>\mro+\mpi\sim 4\pi F_\pi$. 
Nevertheless, our results point to a different conclusion than the one 
reached in Ref. \cite{BKM96}. Specifically, we find that because of
$V'\pi$ structure in the continuum, the $3\pi$ state could play
a non-trivial role in the isoscalar EM channel -- even apart from the 
effect of pure $3\pi\leftrightarrow V$ resonances. Regarding the
strangeness channel, we also note that
the $3\pi$-current vertex employed in Ref. \cite{BKM96}, 
derived from the Wess-Zumino-Witten term, 
gives no hint of any OZI-violating hadronic structure effects which 
would generate a non-zero $\bra{3\pi}\bar{s}\gamma_\mu s\ket{0}$ matrix 
element. To ${\cal O}(q^7)$
in CHPT, then, the un-correlated $3\pi$ continuum does not contribute to
the $F_i^{(s)}(t)$. In order to find such a contribution,
we have relied on hadron phenomenology, including the observation of the 
OZI-violating $\phi\to\rho\pi$ decay and the presence of a $\rho\pi$
resonance in the $p\bar{p}\to 3\pi$ reaction. 

With respect to the pole analyses of Refs. 
\cite{MuI96,Jaf89,HaM96,For96}, we have found
evidence of non-negligible $\phi$ resonance contribution to the
$F_i^{(s)}(t)$ without relying exclusively on the
validity of the pole approximation for the $F_i^{(a)}(t)$
or on the presence of a strong $\phi NN$ coupling obtained from pole
model fits to the $F_i^{I=0}(t)$. At
this time, however, we are unable to determine whether the pure pole
approximation effectively and accurately includes the effect of the
$\phi$ as it arises via a $\rho\pi$ resonance in the $3\pi$ continuum
as well as in the $K\bar{K}$ continuum as analyzed in Ref. 
\cite{MuH97,Mus97}. Within the framework employed
here, neither the $3\pi\leftrightarrow\rho\pi\leftrightarrow\phi$ 
effect, nor the $K\bar{K}\leftrightarrow\phi$ contribution, 
approaches in magnitude the predictions of Refs.
\cite{MuI96,Jaf89,HaM96,For96} 
for the strangeness radius. The corresponding predictions for the strange
magnetic moment, however, are commensurate. An
independent theoretical confirmation of the pole model predictions would
require a more sophisticated treatment of the $N\bar{N}\to 3\pi$, 
$N\bar{N}\to K\bar{K}$, {\em etc.} amplitudes than undertaken here,
including the effects of resonant and non-resonant multi-meson rescattering.
Indeed, studies of $KN$ scattering amplitudes \cite{Mus97}, 
as well as theoretical
models for $N\bar{N}\to\rho\pi$ \cite{Mul95}, suggest that rescattering 
corrections could significantly modify the spectral functions employed in 
the present calculation.

We make no pretense of having performed a definitive analysis. Rather,
within the level of approximation employed in many model calculations,
we have simply demonstrated the prospective importance of contributions 
to the strangeness form factors arising from mesonic intermediate
states containing no valence $s$ or $\bar{s}$-quarks. Although a
na\"\i ve interpretation of the OZI-rule suggests that such contributions
ought to be negligible compared to those from strange intermediate states,
we conclude
that realistic treatments of the $F_i^{(s)}(t)$, using effective hadronic
approaches, ought to take the former contributions into consideration. 
Our results also agree with the conclusions of Ref. \cite{Mei97},
suggesting that a more careful treatment of the $3\pi$ continuum
in the isoscalar EM form factors is warranted.

\section*{Acknowledgement}
We wish to thank R.L. Jaffe, N. Isgur, U.-G. Mei{\ss}ner and D. Drechsel 
for useful discussions. HWH has been supported by 
the German Academic Exchange Service (Doktorandenstipendien HSP III). 
MJR-M has been supported
in part under U.S. Department of Energy contract \# DE-FG06-90ER40561
and under a National Science Foundation Young Investigator Award.

\newpage
\begin{center}
{\Large\bf Figures}
\end{center}
\medskip

\vspace{6cm}

\begin{thebibliography}{99}
\bibitem{comm}We note the study of Ref. \protect\cite{Mei97}, in which the 
              effect of correlated $\rho\pi$ exchange was included in the 
              isoscalar EM spectral functions. The authors find a significant
              reduction in the value of $g_\sst{\phi NN}$ from the pure pole
              analyses. In contrast to the present study, however, they
              assumed that the $\rho\pi$ state does not couple to the
              strangeness vector current.
\end{thebibliography}
\end{document}